\documentclass[12pt,preprint]{aastex}

\slugcomment{ApJ Letters, accepted 5 July 2007}
\shorttitle{Distance and Environment of HESS J1023-575} 
\shortauthors{T. M.\ Dame}

\newcommand{\etal}{{\rm et al.}}

\newcommand{\kms}{km s$^{-1}$}
\newcommand{\msun}{M$_\odot$}

\begin{document}

\title{On the Distance and Molecular Environment \\
of Westerlund~2 and HESS J1023-575 } 
\author{T. M.\ Dame}
\affil{Harvard-Smithsonian Center for Astrophysics, 60 Garden Street, Cambridge MA 02138 \\
tdame@cfa.harvard.edu}

\begin{abstract} 
The extended TeV gamma-ray source HESS J1023-575 is coincident with the 
massive, young stellar cluster Westerlund~2 (Wd2) and its surrounding HII region RCW 49. 
On the basis of an analysis of the CO emission and 21~cm absorption along the line of sight to Wd2, 
it is argued that this cluster, and by assumption the TeV source as well, must be associated 
with a giant molecular cloud in the far side of the Carina arm with a mass 
of $7.5$ x $10^{5}$ \msun. Analysis of the spatial and velocity structure of the cloud reveals clear 
evidence of interaction with Wd2. The cloud's kinematic distance of 
$6.0 \pm 1.0$ kpc is shown to be consistent with distances inferred both from the 
radius-linewidth relation of molecular clouds and from the foreground gas 
column derived from 230 X-ray sources in Wd2. 
\end{abstract}


  \section{Introduction}
  \label{sec:intro}
In just the past few years, a new generation of imaging atmospheric Cherenkov telescopes, led by the HESS array, have increased the number of confirmed very-high-energy (TeV) gamma-ray sources in the Galaxy from just a few \citep[e.g.,][]{Weekes03} to more than two dozen.  Such observations provide insights into where and how highly relativistic particles are accelerated and may ultimately identify the main sources of Galactic cosmic rays. Some of the Galactic TeV sources can be associated with well-known supernova remnants, long considered the most likely cosmic-ray accelerators, and others with pulsars and X-ray binaries, while some cannot as yet be associated with any known objects. 

Unique among the TeV sources so far identified is HESS J1023-575 \citep{Aharonian07}, an extended source that lies toward Westerlund~2 (hereafter Wd2), a massive stellar cluster containing at least a dozen early O stars and two Wolf-Rayet stars. TeV sources and such massive clusters are both so rare in the Galaxy and their coincidence in this case so close ($\sim6'$) that an association between the two is almost beyond doubt. The TeV source J2032+4130 may be associated with a cluster as well \citep[Cyg OB2;][]{Aharonian05}, but the Cygnus region is so complicated that this association is significantly less secure than in the present case \citep{Butt03, Butt07}. Various models have already been proposed for the TeV emission in active stellar regions such as Wd2 \citep[e.g.,][]{Bykov01, Aharonian07, Bednarek07}, most of them involving turbulent particle acceleration driven by powerful stellar winds, multiple supernovae, or both.

A crucial measure for distinguishing among models is the distance to the TeV source, which hereafter will be assumed to be the same as that to Wd2. The distance to Wd2, however, has a long history of uncertainty, and estimates are not converging with time. Two of the most recent distance determinations are among the most discrepant: \citet{Ascenso07} used deep near-infrared imaging and photometry along with pre-main sequence model tracks to infer a distance of 2.8 kpc, whereas \citet{Rauw07} used spectro-photometric measurements of 12 O stars in the cluster to estimate a distance of $8.3 \pm 1.6$ kpc.  Most previous estimates fall between these values. Thorough reviews of past distance determinations were recently given by \citet{Churchwell04} and \citet{Aharonian07}.

Giant molecular clouds (GMCs) have long provided crucial information about distances for a wide range of Galactic objects.  GMCs contain a large fraction of the Galactic gas and dust, they are generally well defined, are known to be associated with, and often interact with, all the products of star formation, and, most importantly, their velocities can readily be measured with high precision with CO observations.  Here classic techniques in Galactic radio astronomy are used in conjunction with CO and 21~cm surveys to argue that Wd2 and its TeV source are almost certainly associated with a very large GMC lying in the far side of the Carina spiral arm at a distance of $6.0 \pm 1.0$ kpc. This view will be supported by examination of 21~cm absorption, the foreground extinction, the radius-linewidth relation for GMCs, and evidence of interaction between Wd2 and the cloud. 
%

\section{Associated Molecular Cloud}

A reason often cited for difficulty in determining the distance of Wd2 is its direction toward the tangent of the Carina spiral arm.  However, as Figure~1 shows, Wd2 is actually displaced several degrees inward from the tangent, in a direction where the near and far sides of the arm are well separated in velocity from each other and from material near the solar circle at $0$ \kms.  Figure~1 also shows that the most conspicuous molecular concentration along the line of sight to Wd2 lies squarely on the locus of the far Carina arm at a velocity of $\sim11$ \kms.  Here this molecular cloud is called MC8, because it corresponds to cloud No. 8 in the Carina arm cloud catalogue of Grabelsky et al. (1988).

An association between Wd2 and MC8 is strongly implied by an examination of 21~cm absorption against the bright radio continuum of RCW~49, the large HII region surrounding Wd2.  Following the technique described by \citet{Griffiths01}, the average 21~cm absorption spectrum over the central region of RCW~49 was determined and in Figure~2 is compared with the CO emission averaged over roughly the same area.  It is clear from this comparison that all the clouds in the near side of the Carina arm, at negative velocities, as well as the cloud at $+4$ \kms, must lie in front of RCW~49, since corresponding 21~cm absorption features are detected.  Indeed, RCW~49 must lie between the foreground cloud F, which produces strong 21~cm absorption, and the background cloud B, which produces none. A likely systemic velocity for RCW~49 then is halfway between the velocities of these two clouds, which is very close to that of MC8. 

An obvious question raised by Figure~2 is why the strong emission from MC8 itself is not evident in the CO spectrum near $11$ \kms.  The answer is provided in Figure~3, which displays the emission structure of the cloud both spatially and as a function of latitude and velocity.  Figure~3b shows that the velocity of MC8 is fairly constant at $\sim11$ \kms, with a gradient of just a few \kms from bottom to top. In the general direction of RCW~49, however, molecular gas is missing near the systemic velocity of the cloud, and instead clumps are seen lying at higher and lower velocities and apparently connected to the main cloud by weaker emission. These clumps produce the foreground CO line labeled F in Figure~2 and the background CO line labeled B. Figure~3b suggests that F and B may arise from velocity-perturbed clumps in the {\em same cloud}, MC8.  

The energy required to accelerate these clumps is very modest as compared with that available from the known O and W-R stars in Wd2 \citep{Rauw07}.  Assuming N(H$_{2}$)/W$_{CO}$ =  1.8 x 10$^{20}$ cm$^{-2}$ K$^{-1}$ km$^{-1}$ s \citep{Dame01} and a distance of $6.0$ kpc (see below), each clump has a mass of a few 10$^{4}$ \msun, a velocity of $\sim6$ \kms with respect to the main cloud, and a kinetic energy of $\sim2$ x 10$^{49}$ ergs. This value is $\sim3$ orders of magnitude less than the wind output of the W-R and O stars over the cluster's estimated lifetime of $\sim2$ Myr, and one or more past SNe, now extinct, may have contributed energy as well.

The case for identifying clump B as a velocity-perturbed fragment of MC8 is somewhat stronger than that for clump F. As Figure~1 shows, emission at the velocity of clump F ($\sim4$ \kms) is not uncommon in the general direction of Wd2, so that a chance alignment of F with Wd2 is possible.  However, emission peaking above 14 \kms is rare in this general direction and is seen only toward Wd2 (directly below the Wd2 arrow in Fig. 1). Further, the fact that Wd2 and its HII region are optically visible suggests that they lie on the near side of MC8, where a general expansion away from Wd2 would lead primarily to a positive perturbation of the gas velocities, as is observed for clump B. This molecular clump is the one most likely to be closest to Wd2 and the TeV source, perhaps as close as the core radius of the HII region, which is $\sim5 \arcmin$ \citep{Whiteoak97}, or ~9 pc at 6 kpc. Its molecular mass is 7 x $10^4$ \msun, mean radius 23 pc, and mean density 30 H$_{2}$ cm$^{-3}$. 

It's worth noting that the proposed association between Wd2 and MC8 is not contingent on both F and B being velocity-perturbed fragments of this cloud. The mere appearance of a perturbed spatial and velocity structure for MC8 in the direction of Wd2, coupled with the tight velocity constraints imposed by the 21~cm absorption spectrum, provide strong evidence for association. CO observations with much higher angular resolution will be required to determine the true nature of clumps F and B and to investigate in detail the interaction between Wd2 and MC8.  

Calculating the H$_{2}$ mass as above, the total H$_{2}$ mass of MC8 is found to be 7.5 x $10^5$ \msun. This is more than twice the mass of the entire Orion molecular complex \citep{Wilson05} and is similar to that of the most massive complexes in the inner Galaxy \citep{Dame86}. It is not surprising that a cluster as rich in early-type stars as Wd2 is found in one of the Galaxy's more massive molecular complexes. 

%

\section{Distance}

Determining the distance to MC8 is in principle straightforward, because the positive velocity of MC8 places it beyond the solar circle, where, on the assumption of pure circular rotation, there is a monotonic relationship between velocity and distance. However, the Galactic rotation curve is more uncertain beyond the solar circle than within, and one must consider, in addition, the uncertainty imposed by random and systematic  non-circular motions. Given that the well-determined rotation curve inside the solar circle is flat to within a few \kms kpc$^{-1}$ \citep{Brand93} and that the rotation curves of external spirals are generally flat or slowly rising to well beyond the solar radius \citep{Binney98}, it is very unlikely that the rotation curve between the solar circle and MC8 has a significant slope. Using more than 200 HII regions and reflection nebulae with spectro-photometric distances, \citet{Brand93} derived a slope of $\sim1$ \kms  kpc$^{-1}$ for the rotation curve beyond the solar circle. Use of this curve with R$_\odot$ = 8.5 kpc and a velocity of 11 \kms for MC8 implies a distance of 6.0 kpc.  Changing the slope of the rotation curve of the outer Galaxy by an improbably large amount with respect to the Brand \& Blitz curve, $\pm5$ \kms kpc$^{-1}$, changes the derived distance by only  $\pm0.35$ kpc. 
	
In considering the effects of random and systematic noncircular motions, it is important to keep in mind that the Galaxy appears overall to be a very orderly system. The smooth and well-defined loci of terminal velocities with longitude in the first and fourth quadrants, and their agreement between quadrants to better than 10 \kms \citep{Blitz91}, tightly constrain the amplitudes of noncircular motions. Random jitter about these smooth curves implies a one-dimensional velocity dispersion of $\sim4.2$ \kms for molecular clouds \citep{Stark89}, and broad ``bumps'' on the curves at the spiral arm tangents imply an amplitude of $\sim7$ \kms  for density-wave streaming along the arms \citep{Binney98}. Larger velocity perturbations may occur perpendicular to the arms because of spiral shocks \citep{Roberts69}, but at MC8 a shock perturbation would be directed at a large angle with respect to our line of sight, $\sim74^{\circ}$ assuming an arm inclination of $10^{\circ}$ \citep{Grabelsky88}. The present case is in contrast to the seminal study by \citet{Roberts72} of shocks in the Perseus arm in the second quadrant, where the proposed shocks of amplitude $\sim20$ \kms are directed almost exactly toward the Sun. A similar shock at the position of MC8 in the Carina arm would perturb its velocity along the line of sight by at most 5 \kms. The actual velocity perturbations due to streaming and shocks depend on the exact location of MC8 with respect to the potential minimum of the Carina arm, which is unknown, but the perturbations should be smaller than the full streaming and shock amplitudes. Therefore, random motions, streaming, and shocks together imply a velocity uncertainty for the cloud of $\sim8$ \kms  and a corresponding distance uncertainty of $\pm$0.8 kpc; this increases to $\pm$0.9 kpc when the uncertainty of the rotation curve in the outer Galaxy is included. Thus one can conclude that MC8 along with Wd2, RCW~49, and HESS J1023-575 all lie at a distance of 6.0 kpc with a conservative uncertainty of $\pm$1.0 kpc. 

A rough independent check on the distance to MC8 can be obtained from the radius-linewidth relation derived for Galactic GMCs by \citet{Dame86}. Following this study, the cloud's angular radius is calculated as $(A/\pi)^{1/2}$, where A is the projected area of the cloud as in Figure~3a. Excluding the perturbed section of the cloud toward Wd2 (see Fig. 3b), one obtains a composite cloud linewidth of 8.3 \kms and, at 6.0 kpc, a linear radius of 50 pc. As Figure~4 shows, a distance of 6.0 kpc places Wd2, perhaps somewhat fortuitously, almost exactly on the fit line. Distances in the range 4-8 kpc are allowed by the fairly large scatter.

An independent check on the distance to Wd2 can be obtained by comparing its foreground gas column as derived from X-ray absorption to that derived from the gas tracers.  Very recently, \citet{Tsujimoto07} identified 468 X-ray sources in Wd2 with Chandra and fit each of the stronger 230 of these with an optically thin thermal plasma (APEC) model to derive foreground N(H).  The sources are spread over a cluster core radius of $6\arcmin$ to $7\arcmin$, a good match to the $8.4\arcmin$ resolution of the CO data used here. The distribution of the derived N(H) is strongly peaked at $1.5 \pm 0.5$ x $10^{22}$ cm$^{-2}$ (see their Fig. 5). As discussed above, the 21~cm absorption spectrum (Fig. 3) demonstrates that all the CO and 21~cm emission at v $< 10$ \kms must lie in front of Wd2.  Integrating the nearest CO spectrum to Wd2 over velocities of less than 10 \kms yields N(H$_{2}$) = 0.44 x 10$^{22}$ cm$^{-2}$.  Similarly, integrating the nearest 21~cm spectrum from the HI survey of \citet{Kalberla05}, one obtains N(H) = 0.56 x 10$^{22}$ cm$^{-2}$.  A mean molecular weight per H of 1.36 implies a total foreground N(H) of 1.9 x 10$^{22}$ cm$^{-2}$, in reasonable agreement with the value derived from the X-rays. 

Although it is not possible here to discuss all previous distance estimates for Wd2, it is worth noting that the lowest estimate frequently cited \citep[2.2 kpc;][]{Brand93} actually derives from an unpublished spectro-photometric study in the thesis by \citet{Brand86}.  Another group of low estimates, of order 4 kpc, are kinematic distances based on radio recombination-line measurements of the velocity of RCW~49, which are $\sim0$ \kms \citep[e.g.,][]{Wilson70, Benaglia05}. It has long been known that optically visible HII regions such as RCW~49 tend to be blueshifted with respect to their parent clouds in a manner consistent with the blister model of \citet{Israel78}. Given the evidence of interaction between RCW~49 and MC8 presented above, it seems likely that the relatively low-mass ionized gas will be perturbed toward us by an amount larger than the 6 \kms perturbation of molecular clump B away from us.  Such a blister flow would shift lower both the recombination-line velocity of RCW~49 and its derived kinematic distance. 


  \section{Summary}

The HESS collaboration has recently identified a new type of very-high-energy gamma-ray source, the massive young stellar cluster Wd2. Attempts to understand the extended TeV emission from this complicated region have so far been hampered by a long history of discrepant distance estimates for Wd2, with recent values ranging from 2 to 8 kpc.

Comparison of a CO emission spectrum toward Wd2 with a 21~cm absorption spectrum measured against its surrounding HII region RCW 49 indicates that Wd2 must lie behind a molecular concentration at +4 \kms and in front of one at +16 \kms.  The only substantial molecular concentration in this velocity range is also the most massive cloud along the entire line of sight, a cloud here called MC8.  The Wd2 cluster is so massive and young that it would be expected to lie very near a large reservoir of molecular gas such as MC8. In the direction of Wd2, MC8 appears to be substantially disrupted both spatially and in velocity. One or both of the molecular concentrations at +4 and +16 \kms might be velocity-perturbed fragments of MC8. 

At a velocity of 11 \kms, MC8 lies along the longitude-velocity locus of the far Carina arm, at a kinematic distance of 6.0 kpc.  Uncertainty about the rotation curve in the outer Galaxy and possible deviations from pure circular motion due to density-wave streaming, spiral shocks, and random cloud motions cumulatively imposed an uncertainty of $\pm1.0$ kpc on the kinematic distance. At 6 kpc, the radius and linewidth of MC8 are in good agreement with the radius-linewidth relation derived for large inner-Galaxy clouds, and the foreground extinction derived from the gas tracers is in good agreement with that derived from 230 X-ray sources in Wd2. 


\acknowledgments 
I am grateful to Naomi McClure-Griffiths for data and advice on the Southern Galactic Plane HI survey, Paula Benaglia for data and useful discussions on radio recombination lines, and Yousaf Butt for bringing the discovery of HESS J1023-575 to my attention. 

%

%
%

%

\clearpage

\begin{figure}
\centering
\epsscale{0.8}
\plotone{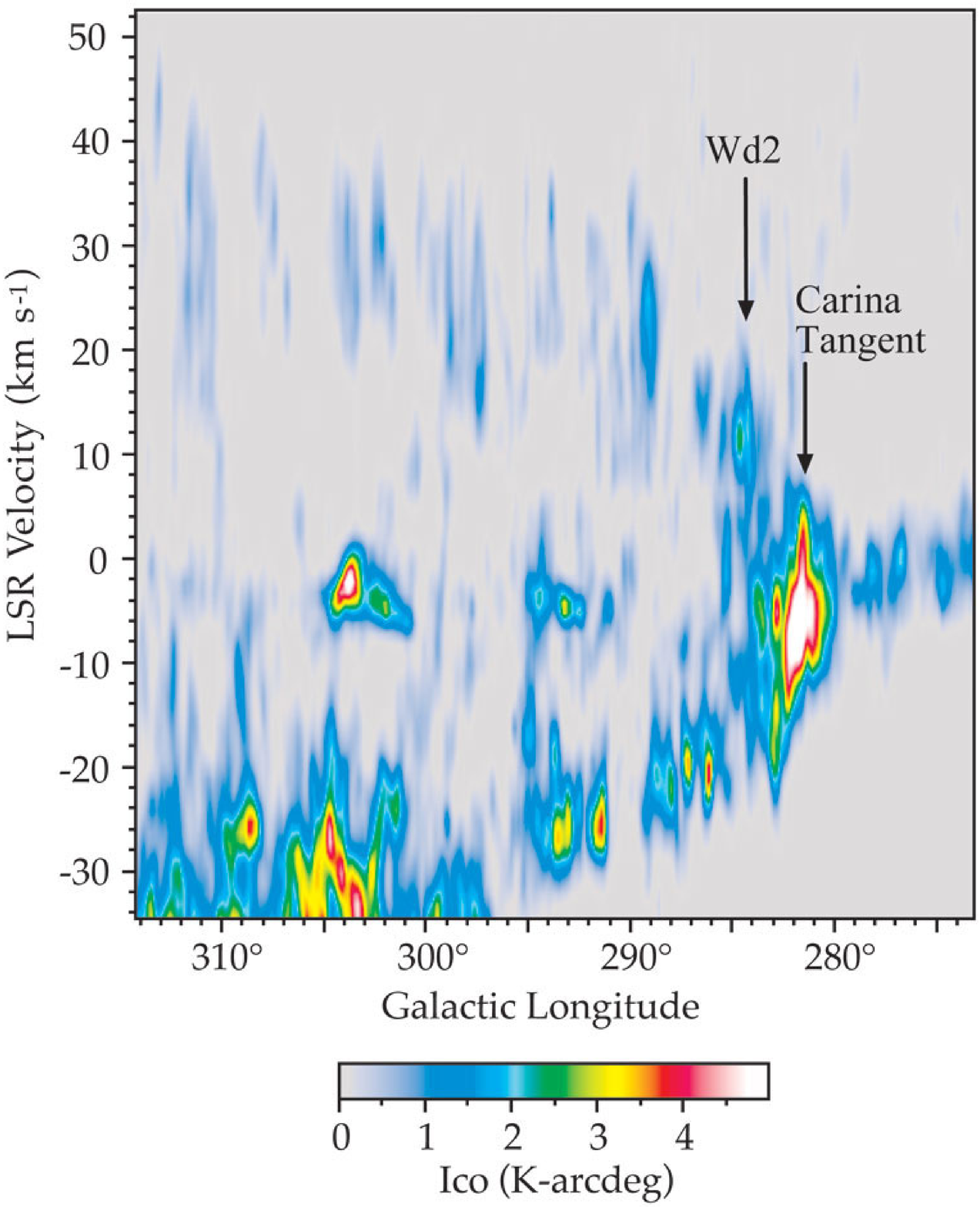}
\caption{Longitude-velocity map of CO emission integrated over a strip $\sim4^{\circ}$ wide in latitude centered on the Galactic plane. The data are from \citet{Dame01}, smoothed slightly to a resolution of $2$ \kms in velocity and $12\arcmin$ in longitude to emphasize the large-scale structure of the Carina arm, which appears as a large loop running from upper left to lower left. Intensities are in units of log(K-arcdeg), ranging from 1 (dark blue) to 5 (white).}
\end{figure}

\begin{figure}
\centering
\epsscale{0.8}
\plotone{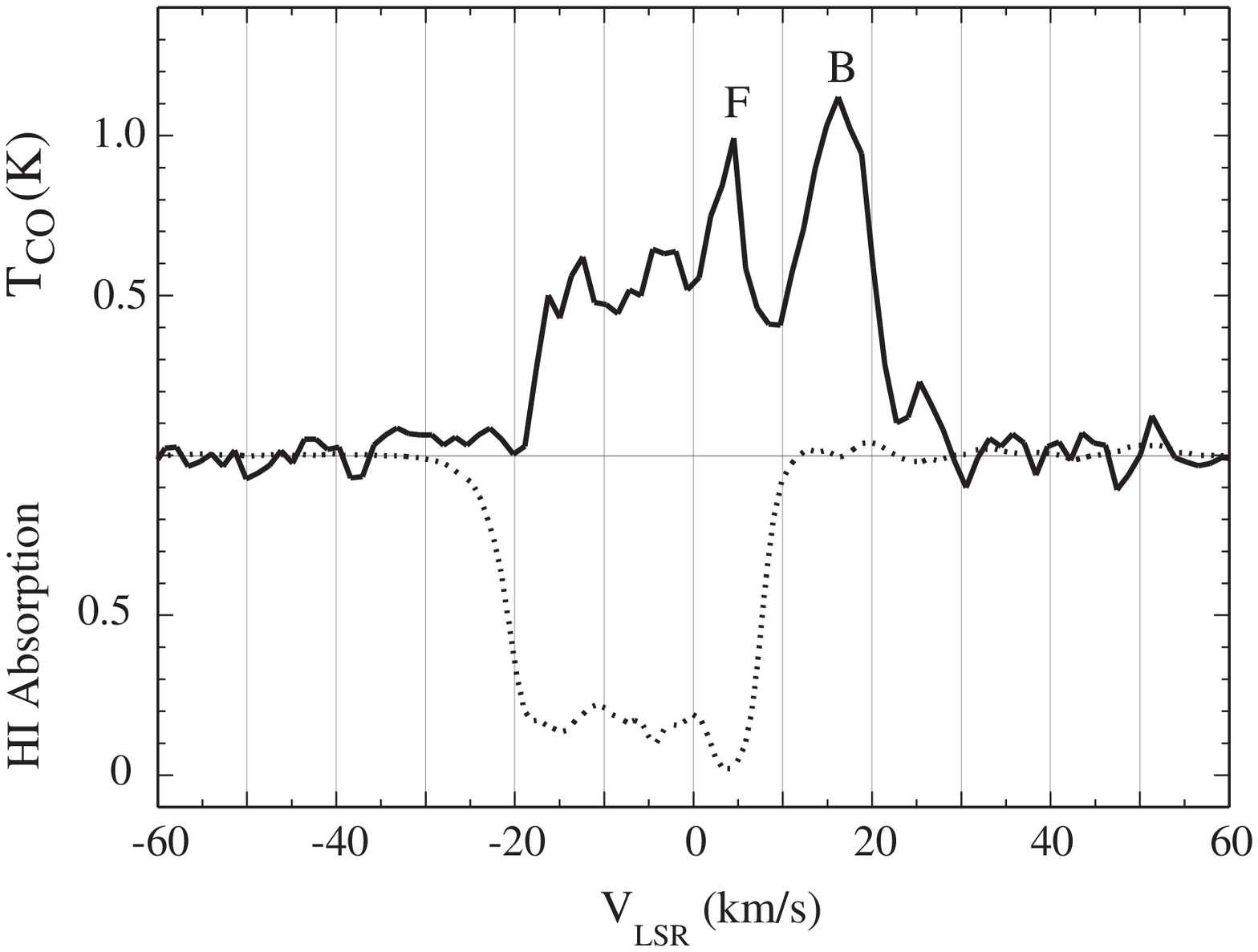}
\caption{Comparison of CO emission  \citep{Grabelsky88} and HI absorption  \citep{Griffiths05} toward the giant HII region RCW~49, which surrounds Wd2. The 21~cm spectra from the brightest part of the continuum emission ($l = 284.3^{\circ}$ to $284.35^{\circ}$, $b = -0.38^{\circ}$ to $Ð0.32^{\circ}$) were averaged as the ``on'' and spectra offset $5\arcmin$ to $10\arcmin$ from the continuum peak were averaged as the ``off''. 
The optical depth spectrum was obtained by subtracting the ``off'' from the ``on'' and dividing by the off-line continuum level of the ``on''. The CO spectra are averaged within $\sim10\arcmin$ of the continuum peak. The letter F marks the highest-velocity line from molecular gas in front of RCW~49, and B marks the lowest-velocity line from molecular gas behind RCW~49.}
\label{fig:mapas}
\end{figure}

\begin{figure*}
\vspace{-6cm}
\centerline{\includegraphics[width=\textwidth]{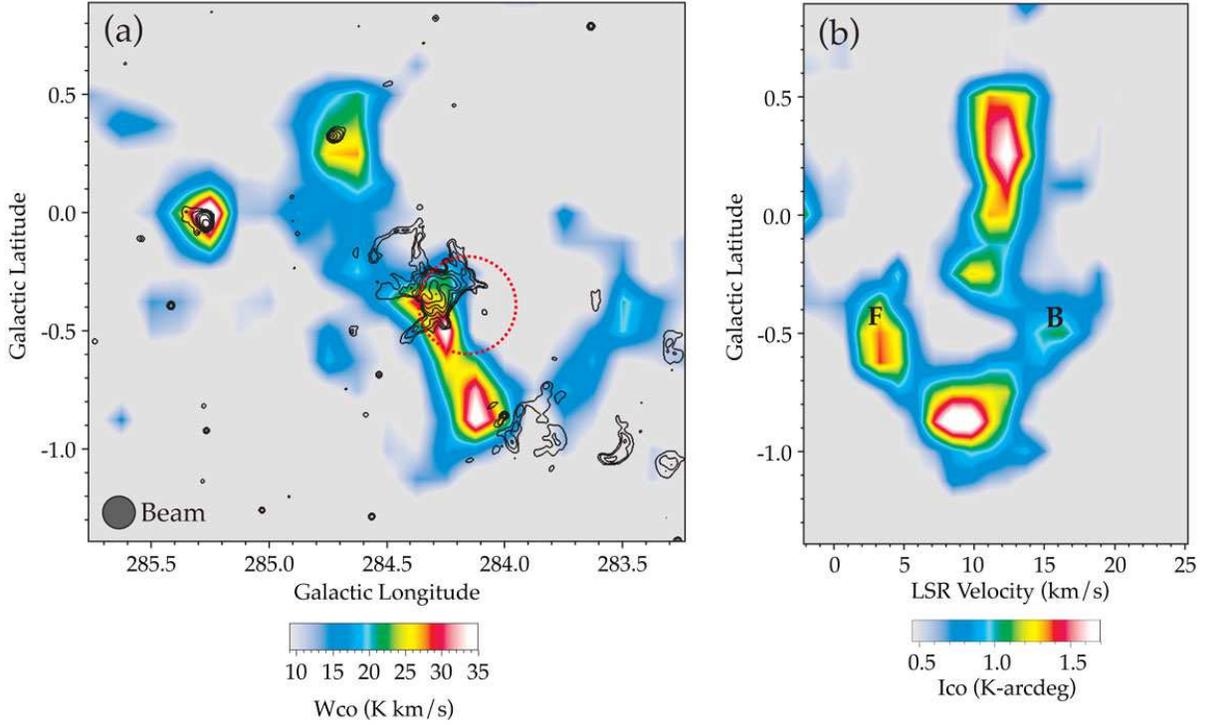}}
\caption{{\em (a)} Color: CO intensity integrated from 0 to 20 \kms in LSR velocity. Contours:  Continuum emission at 843 MHz from the Molonglo Galactic Plane Survey \citep{Green99}; contours are in units of log(mJy beam$^{-1}$), from 1.5 to 3.6 in steps of 0.3. The red dotted circle marks the position and FWHM size of HESS J1023-575. The shell-like continuum source centered near 283.9--0.9 is RCW~48, which may be related to RCW~49 and MC8. The CO source at 285.25+0 is probably unrelated to MC8.
{\em (b)} Latitude-velocity map of CO intensity integrated over the longitude range of MC8, $l = 284^{\circ}$ to $285^{\circ}$.  The letters F and B mark possible fragments of MC8 expanding away from RCW~49; as Fig. 2 shows, F is seen in absorption against RCW~49, but B is not.  
 }
\label{fig:mapaszoom}
\end{figure*}

\begin{figure}
\centering 
\epsscale{0.8}
\plotone{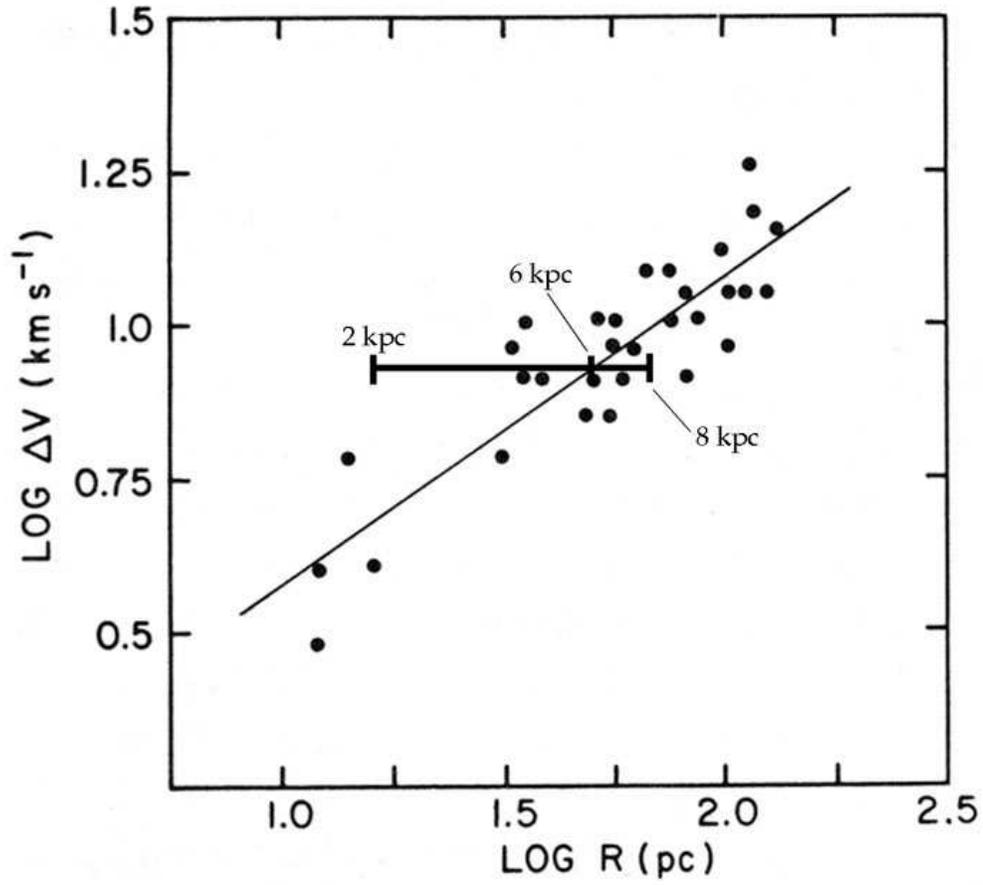}   
\caption{The radius-linewidth relation for giant molecular clouds in the first Galactic quadrant (adapted from Fig. 6 in \citet{Dame86}). The horizontal line marks the locus of possible positions of MC8 for assumed distances ranging from 2 to 8 kpc.  MC8 fits best the radius-linewidth relation at a distance of 6 kpc, the same distance that was derived independently from the cloud's velocity. }
\label{fig:fig4}
\end{figure}

\end{document}